\documentclass[12pt]{iopart}
\newcommand{\dshv}{
\begin{picture}(52,10)
\put(15,3){\line(-5,2){10}}
\put(15,3){\line(-5,-2){10}}
\multiput(15,3)(22,0){2}{\circle*{2}}
\multiput(15,3)(6,0){4}{\line(1,0){4}}
\put(37,3){\line(5,2){10}}
\put(37,3){\line(5,-2){10}}
\end{picture}  }
\newcommand{\zigv}{
\begin{picture}(50,10)
\put(15,3){\line(-5,2){10}}
\put(15,3){\line(-5,-2){10}}
\multiput(15,3)(20.15,0){2}{\circle*{2}}
\multiput(15,3)(0.42,0.42){4}{\circle*{1}}
\multiput(16.26,4.25)(0.5,-0.5){6}{\circle*{1}}
\multiput(18.76,1.75)(0.5,0.5){6}{\circle*{1}}
\multiput(21.26,4.25)(0.5,-0.5){6}{\circle*{1}}
\multiput(23.76,1.75)(0.5,0.5){6}{\circle*{1}}
\multiput(26.26,4.25)(0.5,-0.5){6}{\circle*{1}}
\multiput(28.76,1.75)(0.5,0.5){6}{\circle*{1}}
\multiput(31.26,4.25)(0.5,-0.5){6}{\circle*{1}}
\multiput(33.76,1.75)(0.42,0.42){4}{\circle*{1}}
\put(35.15,3){\line(5,2){10}}
\put(35.15,3){\line(5,-2){10}}
\end{picture}  }
\newcommand{\dotv}{
\begin{picture}(33,10)
\multiput(6,3)(4,0){3}{\circle*{2}}
\put(18,3){\circle*{3}}
\put(18,3){\line(5,2){10}}
\put(18,3){\line(5,-2){10}}
\end{picture}  }
\newcommand{\ddotv}{
\begin{picture}(54,10)
\put(15,3){\line(-5,2){10}}
\put(15,3){\line(-5,-2){10}}
\multiput(15,3)(24,0){2}{\circle*{3}}
\multiput(19,3)(4,0){5}{\circle*{2}}
\put(39,3){\line(5,2){10}}
\put(39,3){\line(5,-2){10}}
\end{picture}  }
\newcommand{\Gausd}{
\begin{picture}(55,10)
\thicklines
\put(6,-1){\line(1,0){40}}
\put(6,1){{\bf k}}
\put(35,1){-{\bf k}}
\end{picture}  }
\newcommand{\dashd}{
\begin{picture}(66,15)
\thicklines
\put(6,-1){\line(1,0){54}}
\put(6,1){{\bf k}}
\put(49,1){-{\bf k}}
\multiput(18,-1)(28,0){2}{\circle*{3}}
\multiput(18,-1)(0.05,0.5){9}{\circle*{1}}
\multiput(46,-1)(-0.05,0.5){9}{\circle*{1}}
\multiput(19,5)(0.3,0.475){9}{\circle*{1}}
\multiput(45,5)(-0.3,0.475){9}{\circle*{1}}
\multiput(23.4,10.6)(0.45,0.25){9}{\circle*{1}}
\multiput(40.6,10.6)(-0.45,0.25){9}{\circle*{1}}
\multiput(30,13.2)(0.25,0){17}{\circle*{1}}
\end{picture}  }
\newcommand{\zigd}{
\begin{picture}(66,15)
\thicklines
\put(6,-1){\line(1,0){54}}
\put(6,1){{\bf k}}
\put(49,1){-{\bf k}}
\multiput(18,-1)(26.6,0){2}{\circle*{3}}
\multiput(18,-1)(-0.1,0.4){11}{\circle*{1}}
\multiput(44.6,-1)(0.1,0.4){11}{\circle*{1}}
\multiput(17,3)(0.4,-0.05){11}{\circle*{1}}
\multiput(45.6,3)(-0.4,-0.05){11}{\circle*{1}}
\multiput(21,2.5)(-0.1,0.4){11}{\circle*{1}}
\multiput(41.6,2.5)(0.1,0.4){11}{\circle*{1}}
\multiput(20,6.5)(0.4,-0.1){11}{\circle*{1}}
\multiput(42.6,6.5)(-0.4,-0.1){11}{\circle*{1}}
\multiput(24,5.5)(0,0.4){11}{\circle*{1}}
\multiput(38.6,5.5)(0,0.4){11}{\circle*{1}}
\multiput(24,9.5)(0.35,-0.22){11}{\circle*{1}}
\multiput(38.6,9.5)(-0.35,-0.22){11}{\circle*{1}}
\multiput(27.5,7.3)(0.15,0.37){11}{\circle*{1}}
\multiput(35.1,7.3)(-0.15,0.37){11}{\circle*{1}}
\multiput(29,11)(0.23,-0.32){11}{\circle*{1}}
\multiput(33.6,11)(-0.23,-0.32){11}{\circle*{1}}
\end{picture}  }
\newcommand{\dotd}{
\begin{picture}(66,15)
\thicklines
\put(6,-1){\line(1,0){54}}
\put(6,1){{\bf k}}
\put(49,1){-{\bf k}}
\multiput(18,-1)(28,0){2}{\circle*{3}}
\multiput(18.5,3)(27,0){2}{\circle*{2}}
\multiput(20.6,7)(22.8,0){2}{\circle*{2}}
\multiput(24,10)(16,0){2}{\circle*{2}}
\multiput(28,12)(8,0){2}{\circle*{2}}
\put(32,12.6){\circle*{2}}
\end{picture}  }
\newcommand{\bldot}{
\begin{picture}(32,12)
\put(9,3.5){\circle{11}}
\multiput(8.8,-1.6)(0.5,0.5){12}{\circle*{1}}
\multiput(9.2,8.6)(-0.5,-0.5){12}{\circle*{1}}
\multiput(6.2,-1)(0.5,0.5){15}{\circle*{1}}
\multiput(11.8,8)(-0.5,-0.5){15}{\circle*{1}}
\multiput(14.5,3.5)(3.5,0){5}{\circle*{2}}
\end{picture}  }
\newcommand{\blddot}{
\begin{picture}(28.5,12)
\put(9,3.5){\circle{11}}
\multiput(8.8,-1.6)(0.5,0.5){12}{\circle*{1}}
\multiput(9.2,8.6)(-0.5,-0.5){12}{\circle*{1}}
\multiput(6.2,-1)(0.5,0.5){15}{\circle*{1}}
\multiput(11.8,8)(-0.5,-0.5){15}{\circle*{1}}
\multiput(14.3,6)(3.5,0.75){4}{\circle*{2}}
\multiput(14.3,1)(3.5,-0.75){4}{\circle*{2}}
\end{picture}  }
\newcommand{\bltdot}{
\begin{picture}(28.5,15)
\put(9,3.5){\circle{11}}
\multiput(8.8,-1.6)(0.5,0.5){12}{\circle*{1}}
\multiput(9.2,8.6)(-0.5,-0.5){12}{\circle*{1}}
\multiput(6.2,-1)(0.5,0.5){15}{\circle*{1}}
\multiput(11.8,8)(-0.5,-0.5){15}{\circle*{1}}
\multiput(14,7)(3.5,0.8){4}{\circle*{2}}
\multiput(14,0)(3.5,-0.8){4}{\circle*{2}}
\multiput(14.5,3.5)(3.5,0){4}{\circle*{2}}
\end{picture}  }
\newcommand{\blzig}{
\begin{picture}(32.5,12)
\put(9,3.5){\circle{11}}
\multiput(8.8,-1.6)(0.5,0.5){12}{\circle*{1}}
\multiput(9.2,8.6)(-0.5,-0.5){12}{\circle*{1}}
\multiput(6.2,-1)(0.5,0.5){15}{\circle*{1}}
\multiput(11.8,8)(-0.5,-0.5){15}{\circle*{1}}
\multiput(14.5,3.5)(0.5,0.5){3}{\circle*{1}}
\multiput(15.5,4.5)(0.5,-0.5){5}{\circle*{1}}
\multiput(17.5,2.5)(0.5,0.5){5}{\circle*{1}}
\multiput(19.5,4.5)(0.5,-0.5){5}{\circle*{1}}
\multiput(21.5,2.5)(0.5,0.5){5}{\circle*{1}}
\multiput(23.5,4.5)(0.5,-0.5){5}{\circle*{1}}
\multiput(25.5,2.5)(0.5,0.5){5}{\circle*{1}}
\multiput(27.5,4.5)(0.5,-0.5){5}{\circle*{1}}
\end{picture}  }
\newcommand{\bldzig}{
\begin{picture}(30,12)
\put(9,3.5){\circle{11}}
\multiput(8.8,-1.6)(0.5,0.5){12}{\circle*{1}}
\multiput(9.2,8.6)(-0.5,-0.5){12}{\circle*{1}}
\multiput(6.2,-1)(0.5,0.5){15}{\circle*{1}}
\multiput(11.8,8)(-0.5,-0.5){15}{\circle*{1}}
\multiput(14.2,5.5)(0.4,0.55){5}{\circle*{1}}
\multiput(15.8,7.7)(0.55,-0.4){5}{\circle*{1}}
\multiput(18,6.1)(0.4,0.55){5}{\circle*{1}}
\multiput(19.6,8.3)(0.55,-0.4){5}{\circle*{1}}
\multiput(21.8,6.7)(0.4,0.55){5}{\circle*{1}}
\multiput(23.4,8.9)(0.55,-0.4){5}{\circle*{1}}
\multiput(25.6,7.3)(0.32,0.44){4}{\circle*{1}}
\multiput(14.2,1.5)(0.4,-0.55){5}{\circle*{1}}
\multiput(15.8,-0.7)(0.55,0.4){5}{\circle*{1}}
\multiput(18,0.9)(0.4,-0.55){5}{\circle*{1}}
\multiput(19.6,-1.3)(0.55,0.4){5}{\circle*{1}}
\multiput(21.8,0.3)(0.4,-0.55){5}{\circle*{1}}
\multiput(23.4,-1.9)(0.55,0.4){5}{\circle*{1}}
\multiput(25.6,-0.3)(0.32,-0.44){4}{\circle*{1}}
\end{picture}  }
\newcommand{\bltzig}{
\begin{picture}(30,15)
\put(9,3.5){\circle{11}}
\multiput(8.8,-1.6)(0.5,0.5){12}{\circle*{1}}
\multiput(9.2,8.6)(-0.5,-0.5){12}{\circle*{1}}
\multiput(6.2,-1)(0.5,0.5){15}{\circle*{1}}
\multiput(11.8,8)(-0.5,-0.5){15}{\circle*{1}}
\multiput(13.7,7)(0.4,0.55){5}{\circle*{1}}
\multiput(15.3,9.2)(0.55,-0.4){5}{\circle*{1}}
\multiput(17.5,7.6)(0.4,0.55){5}{\circle*{1}}
\multiput(19.1,9.8)(0.55,-0.4){5}{\circle*{1}}
\multiput(21.3,8.2)(0.4,0.55){5}{\circle*{1}}
\multiput(22.9,10.4)(0.55,-0.4){5}{\circle*{1}}
\multiput(25.1,8.8)(0.32,0.44){4}{\circle*{1}}
\multiput(13.7,0)(0.4,-0.55){5}{\circle*{1}}
\multiput(15.3,-2.2)(0.55,0.4){5}{\circle*{1}}
\multiput(17.5,-0.6)(0.4,-0.55){5}{\circle*{1}}
\multiput(19.1,-2.8)(0.55,0.4){5}{\circle*{1}}
\multiput(21.3,-1.2)(0.4,-0.55){5}{\circle*{1}}
\multiput(22.9,-3.4)(0.55,0.4){5}{\circle*{1}}
\multiput(25.1,-1.8)(0.32,-0.44){4}{\circle*{1}}
\multiput(14.5,3.5)(0.5,0.5){3}{\circle*{1}}
\multiput(15.5,4.5)(0.5,-0.5){5}{\circle*{1}}
\multiput(17.5,2.5)(0.5,0.5){5}{\circle*{1}}
\multiput(19.5,4.5)(0.5,-0.5){5}{\circle*{1}}
\multiput(21.5,2.5)(0.5,0.5){5}{\circle*{1}}
\multiput(23.5,4.5)(0.5,-0.5){5}{\circle*{1}}
\multiput(25.5,2.5)(0.5,0.5){3}{\circle*{1}}
\end{picture}  }
\begin{document}

\title[]{Critical behavior of n-vector model with quenched
randomness}

\author{J.~Kaupu\v{z}s}

\address{\ Institute of Mathematics and Computer Science,
University of Latvia \\ Rainja Boulevard 29, LV-1459 Riga, Latvia
\\  e--mail: \; kaupuzs@latnet.lv}

\begin{abstract}
We consider the Ginzburg--Landau phase transition model with $O(n)$
symmetry (i.e., the $n$--vector model) which includes a quenched randomness,
i.e., a random temperature disorder.~We have proven rigorously that within
the diagrammatic perturbation theory the quenched randomness does not change
the critical exponents at $n \to 0$, which is in contrast to the
conventional point of view based on the perturbative renormalization group
theory.
\end{abstract}




\section{Introduction}
The phase transitions and critical phenomena is one of the most widely
investigated topic in modern physics. Nevertheless, an eliminated number
of exact and rigorous results are available, and they refer mainly to
the two--dimensional systems~\cite{Baxter} and fractals~\cite{ReMa}.
Rigorous results have been obtained also in four
dimensions~\cite{HaTa} based on an exact renormalization group (RG)
technique~\cite{GaKu}. The RG method, obviously, provides exact results
at $d>4$ (where $d$ is the spatial dimensionality), but
this case is somewhat trivial in view of critical phenomena. In three
dimensions, approximate methods are usually used based on perturbation
theory.

  Here we present a particular result obtained within the diagrammatic
perturbation theory. The Ginzburg--Landau phase transition model with
$O(n)$ symmetry (i.e., the  $n$--vector model) is considered, which
includes a quenched random temperature disorder. The usual prediction
of the perturbative RG field theory~\cite{Wilson,Ma,Justin} is that, in
the case of the spatial dimensionality $d<4$ and small enough $n$
(at  $n=1$ and $n \to 0$, in particular), the critical behavior
of the $n$--component vector model is essentially changed by the quenched
randomness. Here we challenge this conventional point of view based on
a mathematical proof. We have proven rigorously that within the
diagrammatic perturbation theory the critical exponents in the actually
considered model cannot be changed by the quenched randomness at
$n \to 0$.

\section{The model}
   We consider a model with the Ginzburg--Landau Hamiltonian
\begin{eqnarray} \label{eq:Hr}
H/T = \int \left[ \left(r_0+ \sqrt{u} \, f({\bf x}) \right)
\varphi^2({\bf x}) + c \, (\nabla \varphi({\bf x}) )^2 \right] d{\bf x}
\nonumber \\
+ \, uV^{-1} \sum\limits_{i,j,{\bf k}_1,{\bf k}_2,{\bf k}_3 }
\varphi_i({\bf k}_1) \varphi_i({\bf k}_2) \, u_{{\bf k}_1+{\bf k}_2} \,
\varphi_j({\bf k}_3) \varphi_j(-{\bf k}_1-{\bf k}_2-{\bf k}_3)
\end{eqnarray}
which includes a random temperature (or random mass) disorder
represented by the term $\sqrt{u}\,f({\bf x})\,\varphi^2({\bf x})$.
For convenience, we call this model the random model. In Eq.~(\ref{eq:Hr})
$\varphi({\bf x})$ is an $n$--component vector (the order parameter
field) with components
$\varphi_i({\bf x})=V^{-1/2} \sum_{\bf k} \varphi_i({\bf k}) e^{i{\bf kx}}$,
depending on the coordinate ${\bf x}$, and
$f({\bf x})=V^{-1/2} \sum_{\bf k} f_{\bf k} e^{i{\bf kx}}$ is a random
variable with the Fourier components
$f_{\bf k}=V^{-1/2} \int f({\bf x}) e^{-i{\bf kx}} d{\bf x}$.
Here $V$ is the volume of the system, $T$ is the temperature, and
$\varphi_i({\bf k})$ is the Fourier transform of $\varphi_i({\bf x})$.
An upper limit of the magnitude of wave vector $k_0$ is fixed. It means
that the only allowed configurations of the order parameter field are
those corresponding to $\varphi_i({\bf k})=0$ at $k>k_0$. This is the
limiting case $m \to \infty$ ($m$ is integer) of the model where all
configurations of $\varphi({\bf x})$ are allowed, but Hamiltonian
(\ref{eq:Hr}) is completed by term
$\sum\limits_{i,{\bf k}} (k/k_0)^{2m} {\mid \varphi_i({\bf k}) \mid}^2$.

    The perturbation expansions of various physical quantities in powers
of the coupling constant  $u$  are of interest. In this case  $n$  may be
considered as a continuous parameter. In particular, the case
$n \to 0$ has a physical meaning
describing the statistics of polymers~\cite{Justin}.

   The system is characterized by the two--point correlation function
$G_i({\bf k})$ defined by the equation
\begin{equation}
\left< \varphi_i({\bf k}) \varphi_j(-{\bf k}) \right>
= \delta_{i,j} \, G_i({\bf k}) \;.
\end{equation}
Because of the $O(n)$ symmetry of the considered model, we have
$G_i({\bf k}) \equiv G({\bf k})$, i.e., the index  $i$  may be omitted.
It is supposed that the averaging is performed over the $f({\bf x})$
configurations with a Gaussian distribution for the Fourier components
$f_{\bf k}$, i.e., the configuration $ \{ f_{\bf k} \}$ is taken with
the weight function
\begin{equation} \label{eq:P}
P(\{ f_{\bf k} \}) = Z_1^{-1} \exp \left( - \sum\limits_{\bf k}
b({\bf k}) \, {\mid f_{\bf k} \mid}^2 \right) \;,
\end{equation}
where
\begin{equation} \label{eq:Z}
Z_1 = \int \exp \left( - \sum\limits_{\bf k}
b({\bf k}) \, {\mid f_{\bf k} \mid}^2 \right) \, D(f) \;,
\end{equation}
and $b({\bf k})$ is a positively defined function of $k$.
Eq.~(\ref{eq:Hr}) defines the simplest random model considered
in~\cite{Ma} (according to the universality hypothesis, the factor
$\sqrt{u}$ does not make an important difference). Our random
model describes a quenched randomness since the distribution over the
configurations $\{ f_{\bf k} \}$ of the random variable is given
(by Eqs.~(\ref{eq:P}) and (\ref{eq:Z})) and does not depend neither on
temperature nor the configuration
$\{ \varphi_i({\bf k}) \}$ of the order parameter field.
More precisely, the common distribution over the configurations
$\{ f_{\bf k} \}$ and $\{ \varphi_i({\bf k}) \}$ is given by
\begin{equation}
P( \{ f_{\bf k} \} , \{ \varphi_i({\bf k}) \} )
= P( \{ f_{\bf k} \} ) \times Z_2^{-1}( \{ f_{\bf k} \} ) \,
\exp(-H/T) \;,
\end{equation}
where $Z_2( \{ f_{\bf k} \} ) = \int \exp(-H/T) D(\varphi)$ and $H$ is
defined by Eq.~(\ref{eq:Hr}). For comparison, the common distribution
is the equilibrium Gibbs distribution in a case of annealed randomness
sometimes considered in literature.

\section{A basic theorem}
 We have proven the following theorem. \bigskip

{\it Theorem}. \, In the limit $n \to 0$, the perturbation expansion
of the correlation function  $G({\bf k})$ in  $u$  power series for the
random model with the Hamiltonian (\ref{eq:Hr}) is identical to the
perturbation expansion for the corresponding model with the Hamiltonian
\begin{eqnarray} \label{eq:Hp}
H/T = \int \left[ r_0 \, \varphi^2({\bf x})
+ c \, (\nabla \varphi({\bf x}) )^2 \right] d{\bf x} \nonumber \\
+ \, uV^{-1} \sum\limits_{i,j,{\bf k}_1,{\bf k}_2,{\bf k}_3 }
\varphi_i({\bf k}_1) \varphi_i({\bf k}_2) \, \tilde u_{{\bf k}_1+{\bf k}_2} \,
\varphi_j({\bf k}_3) \varphi_j(-{\bf k}_1-{\bf k}_2-{\bf k}_3)
\end{eqnarray}
where $\tilde u_{\bf k}=u_{\bf k}- {1 \over 2}
\left< {\mid f_{\bf k} \mid}^2 \right>$. \medskip

   For convenience, we call the model without the term
$\sqrt{u} \, f({\bf x}) \, \varphi^2({\bf x})$ the pure model, since this
term simulates the effect of random impurities~\cite{Ma}.
\bigskip

 {\it Proof of the theorem}. \, According to the rules of the diagram
technique, the formal expansion for $G({\bf k})$ involves all connected
diagrams with two fixed outer solid lines. In the case of the pure model,
diagrams are constructed of the vertices {\mbox \zigv,} with factor
$-uV^{-1} \tilde u_{\bf k}$ related to any zigzag line with wave vector
${\bf k}$.
The solid lines are related to the correlation function in the Gaussian
approximation $G_0({\bf k})=1/ \left(2r_0+2ck^2 \right)$. Summation over
the components $\varphi_i({\bf k})$ of the vector $\varphi({\bf k})$
yields factor $n$ corresponding to each closed loop of solid lines in the
diagrams. According to this, the formal perturbation expansion is defined
at arbitrary $n$. In the limit $n \to 0$, all diagrams of $G({\bf k})$
vanish except those which do not contain the closed loops. In such a way,
for the pure model we obtain the expansion
\begin{equation} \label{eq:expu}
G({\bf k})= \Gausd + \, \zigd + \, ... \;.
\end{equation}
 In the case of the random model, the diagrams are constructed of
the vertices \dshv and {\mbox \dotv.} Besides, it is important that only
those diagrams give a nonzero contribution where each dotted line is
coupled to another dotted line. The factors
$uV^{-1} \left< {\mid f_{\bf k} \mid}^2 \right>$ correspond to the
coupled dotted lines and the factors $-uV^{-1} u_{\bf k}$ correspond to
the dashed lines. Thus, we have
\begin{equation} \label{eq:exra}
G({\bf k}) = \Gausd + \left[ \dashd + \dotd \right] + \, ... \;.
\end{equation}
In our notation the combinatorial factor corresponding to any specific
diagram is not given explicitly, but is implied in the diagram itself.
In the random model, first the correlation function $G({\bf k})$
is calculated at a fixed $\{ f_{\bf k} \}$ according to the distribution
$Z_2^{-1} \left( \{ f_{\bf k} \} \right) \, \exp(-H/T)$
(which corresponds to diagrams where solid lines are coupled, but the
dotted lines with factors $-\sqrt{u} \,V^{-1/2} f_{\bf k}$ are not coupled),
performing the averaging with the weight (\ref{eq:P}) over the
configurations of the random variable
(i.e., the coupling of the dotted lines) afterwards.
According to this procedure, the diagrams of the random model in
general (not only at $n \to 0$) do not contain parts like {\mbox \bldot,}
{\mbox \blddot,} {\mbox \bltdot,} etc. Such parts would appear after the
coupling of dotted lines only if unconnected (i.e., consisting of separate
parts) diagrams with fixed $\{f_{\bf k}\}$ would be considered.

     Thus, in the considered random model, the term of the
$l$--th order in the perturbation expansion of $G({\bf k})$
in $u$ power series is represented by diagrams constructed of a number
$M_1$ of vertices \dshv and an even number $M_2$ of vertices \dotv (i.e.,
$M_2/2$ \, double--vertices \ddotv), such that $l=M_1+M_2/2$. In the pure
model, defined by Eq.~(\ref{eq:Hp}), this term is represented by diagrams
constructed of $l$ vertices {\mbox \zigv.} It is clear and evident from
Eqs.~(\ref{eq:expu}) and (\ref{eq:exra}) that all diagrams of the random
model are obtained from those of the pure model if any of the zigzag
lines is replaced either by a dashed or by a dotted line, performing
summation over all such possibilities. Note that such a method is valid in
the limit $n \to 0$, but not in general. The problem is that, except
the case $n \to 0$, the diagrams of the pure model contain parts like
{\mbox \blzig,} {\mbox \bldzig,} {\mbox \bltzig,} etc. If all the depicted
here zigzag lines are replaced by the dotted lines, then we obtain diagrams
which are not allowed in the random model, as it has been explained
before. At $n \to 0$, the only problem is to determine the
combinatorial factors for the diagrams obtained by the above replacements.
For a diagram constructed of $M_1$ vertices \dshv and $M_2$ vertices \dotv
the combinatorial factor is the number of possible different couplings of
lines, corresponding to the given topological picture, divided by
$M_1 ! M_2 !$.

     It is suitable to make some systematic grouping of the diagrams of
the random model. The following consideration is valid not only for
the diagrams of the two--point correlation function, but also for free
energy diagrams (connected diagrams without outer lines) and for the
diagrams of $2m$--point correlation function (i.e., the
diagrams with $2m$ fixed outer solid lines, containing no separate parts
unconnected to these lines). It is supposed that at $n \to 0$ the main terms
are retained, which means that the free energy diagrams contain a single loop
of solid lines. We define that all diagrams which can be obtained from the
$i$--th diagram (i.e., the diagram of the $i$--th topology) of the pure
model, belong to the  $i$--th group. The sum of the diagrams of the
$i$--th group can be found by the following algorithm.

1. First, the $i$--th diagram of pure model is depicted in
an a priori defined way.

2. Each vertex \zigv is replaced either by the vertex {\mbox \dshv,} or
by the double-vertex {\mbox \ddotv,} performing the summation over all
possibilities. Besides, all vertices \dshv and \dotv and all lines are
numbered before coupling, and all the distributions of the numbered
vertices and lines over the numbered positions (arranged according to the
given picture defined in step 1 and according to the actually considered
choice, defining which of the vertices \zigv must be replaced by \dshv
and which of them must be replaced by \ddotv) are counted as different.
Each specific realization is summed over with the weight $1/(M_1!M_2!)$.

3. To ensure that each specific diagram is counted with the correct
combinatorial factor, the result of summation in step 2 is divided by the
number of independent symmetry transformations (including the identical
transformation) $S_i$ for the considered $i$--th diagram
constructed of vertices {\mbox \zigv,} where the symmetry transformation
of a diagram is defined as any possible redistribution (such that the
outer solid lines are fixed) of vertices and coupled lines not changing
the given picture. Really, the coupling of lines is not changed if any of
the symmetry transformations with any of the specific diagrams of the
$i$--th group is performed, whereas, according to the algorithm of
step 2, original and transformed diagrams are counted as different.

  It is convenient to modify step 2 as follows. Choose any one replacement
of the vertices \zigv by \dshv and {\mbox \ddotv,} and perform the summation
over all such possibilities. For any specific choice we consider only one
of the possible $M_1!M_2!$ distributions of the numbered $M_1$ vertices
\dshv and $M_2$ vertices \dotv over the fixed numbered positions, and make
the summation with weight $1$ instead of the summation over
$M_1!M_2!$ equivalent
(i.e., equally contributing) distributions with the weight $1/(M_1!M_2!)$.

   Note that the location of any vertex \dshv is defined by fixing the
position of dashed line, the orientation of which is not fixed. According
to this, the summation over all possible distributions of lines (numbered
before coupling) for one fixed location of vertices (as consistent with
the modified step 2) yields factor $8^{M_1}4^{M_2/2}$. The $i$--th
diagram of the pure model also can be calculated by such an algorithm.
In this case the summation over all
possible line distributions yields a factor of $8^l$, where $l=M_1+M_2/2$\,
is the total number of vertices \zigv in the  $i$--th diagram. Obviously,
the summation of diagrams of the  $i$--th  group can be performed with
factors $8^l$ instead of $8^{M_1}4^{M_2/2}$, but in this case factors
${1 \over 2} uV^{-1} \left< {\mid f_{\bf k} \mid}^2 \right>$
must be related to the coupled dotted lines instead of
$uV^{-1} \left< {\mid f_{\bf k} \mid}^2 \right>$. In this case the
summation over all possibilities where zigzag lines are replaced by dashed
lines with factors $-uV^{-1}u_{\bf k}$ and by dotted lines with factors
${1 \over 2} uV^{-1} \left< {\mid f_{\bf k} \mid}^2 \right>$, obviously,
yields a factor
$uV^{-1} \left(-u_{\bf k}+{1 \over 2} \left< {\mid f_{\bf k} \mid}^2
\right> \right) \equiv -uV^{-1} \tilde u_{\bf k}$ corresponding to each
zigzag line with wave vector ${\bf k}$. Thus, the sum over the diagrams of
the  $i$--th group is identical to the  $i$--th diagram of the pure model
defined by Eq.~(\ref{eq:Hp}). By this the theorem has proved.

\section{Remarks and conclusions}
  The theorem has been formulated for the two--point correlation function,
but the proof, in fact, is more general, as regards the relation between
diagrams of random and pure models. Thus, the statement of the theorem is
true also for free energy and for  $2m$--point  correlation function.

  According to the proven theorem and this remark, at $n \to 0$
the considered pure and random models cannot be distinguished within the
diagrammatic perturbation theory. Thus, if, in principle, critical
exponents can be determined from the diagrammatic perturbation theory at
$n \to 0$, then, in this limit, the critical exponents for the random
model are the same as for the pure model. We think that in reality
correct values of critical exponents can be determined from the
diagrammatic perturbation theory, therefore the quenched random
temperature disorder does
not change the universality class at $n \to 0$. This our conclusion
correlates with results of some other authors. In particular, there is a
good evidence that the universality class is not changed by the quenched
randomness at $n=1$. It has been shown by extensive Monte--Carlo
simulations of two--dimensional dilute Ising models \cite{SzIg} that the
critical exponent of the defect magnetization is a continuous function of
the defect coupling. By analyzing the stability conditions, it has been
concluded in Ref.~\cite{SzIg} that the critical exponent $\nu$ of the
bulk correlation length of the random Ising model does not depend on the
dilution, i.e., $\nu=1$ at $d=2$ both for diluted and not diluted Ising
models. The standard (pertubative) RG method predicts
the change of the universality class by the quenched randomness.
We think, this is a false prediction. The fact that the standard
RG method provides incorrect result is not surprising, since it has
been demonstrated (in fact, proven) in Ref.~\cite{Kaupuzs} that this
method is not valid at $d<4$.

\References
\bibitem[1]{Baxter} Rodney J.~Baxter, {\it Exactly Solved Models in
Statistical Mechanics} (Academic Press, London etc.,1982)
\bibitem[2]{ReMa} Jose Arnaldo Redinz, Aglae C.~N.~de~Magelhaes,
{\it Phys. Rev. B} \, {\bf 51}, 2930 (1995)
\bibitem[3]{HaTa} T.~Hara, H.~Tasaki, {\it J. Stat. Phys.}
{\bf 47}, 99 (1987)
\bibitem[4]{GaKu} K.~Gawedzki, A.~Kupiainen, {\it Commun. Math. Phys.}
{\bf 99}, 197 (1985)
\bibitem[5]{Wilson} K.G.~Wilson, M.E.~Fisher, {\it
Phys.Rev.Lett.} {\bf 28}, 240 (1972)
\bibitem[6]{Ma} Shang--Keng Ma, {\it Modern Theory of Critical
Phenomena} (W.A.~Benjamin, Inc., New York, 1976)
\bibitem[7]{Justin} J.~Zinn--Justin, {\it Quantum Field Theory and
Critical Phenomena} (Clarendon Press, Oxford, 1996)
\bibitem[8]{SzIg} F.~Szalma and F.~Igloi, {\it Abstracts MECO 24, P69} \\
(March 8-10, Lutterstadt Wittenberg, Germany, 1999)
\bibitem[9]{Kaupuzs} J.~Kaupu\v{z}s, {\it archived as cond-mat/0001414 of
xxx.lanl.gov archive}, January 2000
\endrefs
\end{document}